**Catalytic action of two-dimensional layered materials (WS$_2$, and MoS$_2$) on hydrogen sorption properties of MgH$_2$**

*Satish Kumar Verma[1], Mohammad Abu Shaz[1], Thakur Prasad Yadav[1,2]\**

[1]Hydrogen Energy Centre, Department of Physics, Banaras Hindu University, Varanasi-221005, India.

[2]Department of Physics, Faculty of Science, University of Allahabad, Prayagraj-211002, India.

**Abstract:**

The present study reports the catalytic action of two-dimensional (2D) layered materials (MoS$_2$ and WS$_2$) for improving the de/re-hydrogenation kinetics of MgH$_2$. The MgH$_2$ start desorbing at 277 $^\circ$C with a hydrogen storage capacity of 5.95 wt% in the presence of WS$_2$ catalyst whereas onset desorption temperature of MgH$_2$ catalyzed by MoS$_2$ is 330 $^\circ$C. The MgH$_2$-WS$_2$ absorbed hydrogen ~ 3.72 wt% within 1.3 minutes at 300 $^\circ$C under 13 atm hydrogen pressure and it desorbed ~5.57 wt% within 20 minutes at 300 $^\circ$C under 1 atm hydrogen pressure. We have performed 25 cycles of dehydrogenation (under 1 atm hydrogen pressure at 300 ºC) and re-hydrogenation (under 13 atm hydrogen pressure at 300 °C) to ensure cyclic stability of catalyzed version of MgH$_2$ where MgH$_2$-WS$_2$ shows better cyclic stability than MgH$_2$-MoS$_2$. MgH$_2$-WS$_2$ also shows the lower reaction activation energy ~117 kJ/mol as compare to other catalyzed and uncatalyzed samples. On the other hand, these catalysts (WS$_2$ and MoS$_2$) do not have any impact on the thermodynamical parameters that is change in enthalpy.

**\*Corresponding author Email:** yadavtp@gmail.com

## 1. Introduction

A crucial and promising area of research for onboard hydrogen applications is the development of safe and efficient hydrogen storage. The solid-state approach is one of the most appropriate, secure, and effective ways to store hydrogen among the several methods that can be used, including gaseous, liquid, and solid-state storage [1,2]. Due to its high hydrogen storage capacity (110 g/L volumetric and 7.6 wt% gravimetric), low cost, light weight, and large abundance (in the form of Mg) in earth crust (8[th] most) and seawater (3rd most), $MgH_2$ is a leading choice for hydrogen storage in the solid-state mode [3–6]. According to the United States Department of Energy (US DOE) technical targets for hydrogen storage systems [7], $MgH_2$ has certain advantages that make it a viable option. The high dehydrogenation temperature (above 400 $^\circ$C), slow kinetics (hydrogen de/re-hydrogenation kinetics 0.4 kg-$H_2$/min), and high thermodynamic properties (high reaction enthalpy 74 kJ/mol) of $MgH_2$ prevent it from being a suitable material for onboard applications even with these advantages [8–10]. In recent years, the creation of suitable catalyst(s), alloys, composite materials with complicated hydrides, and scaffolding have all been used as feasible methods to improve the hydrogen storage performance of $MgH_2$ [11,12]. The use of various types of catalysts and additives to enhance the performance of Mg/$MgH_2$ has been the subject of several studies by various research organizations [13–17].

Another application for the 2D materials is as a catalyst for improving the hydrogen characteristics of $MgH_2$ [18–22]. Due to its enormous surface area, ballistic conduction, thermal conductivity, mechanical stability, and light weight, graphene, which has a 2D planer structure with sp$^2$ carbon atoms arranged in a hexagonal framework, has attracted a lot of attention as a catalyst and as a template material for hydrogen storage application in $MgH_2$ [23,24]. $MgH_2$'s de/rehydrogenation kinetics exhibit effective catalytic behavior in the graphene layer, which also inhibits $MgH_2$'s agglomeration and grain growth [3,5,25]. Liu et al., for instance, have created $MgH_2$-5% Gr nanosheets [26]. They have demonstrated that graphene nanosheets offer a significant hydrogen diffusion pathway and prevent $MgH_2$ from aggregating. According to Huang et al., [27] report's $MgH_2$ nanoparticles supported by graphene exhibit remarkable hydrogen sorption kinetics and cyclic stability. Due to the strong interaction between graphene and MgH$_2$ nanoparticles and the prevention of nanoparticle agglomeration, the MgH$_2$ nanoparticles demonstrated excellent hydrogen storage performance. Additionally, grapheme prevents the aggregation of nanoparticles during the rehydrogenation of MgH$_2$, according to a theoretical study using molecular dynamics simulation [28]. Rough studies are still required to determine the impact of graphene and other 2D layered materials on MgH$_2$, even though some prior studies have shown the remarkable catalytic/co-catalytic and agglomeration blocking properties of Gr on MgH$_2$.

We have examined a comparison between WS$_2$ and MoS$_2$ as a catalyst for enhancing hydrogen sorption properties of MgH$_2$. WS$_2$ and MoS$_2$ are suitable alternatives to graphene for the catalytic action on MgH$_2$ due to their high conductivity (metallic nature), thermal stability, and strong catalytic behavior [29,30]. Tungsten (W) and Molybdenum (Mo) are sandwiched between two Sulphur layers with weak Van der Waals interactions in the family of layered transition-metal dichalcogenides (TMDs) materials that include WS$_2$ and MoS$_2$. The re/de-hydrogenation kinetics, and catalytic behavior of WS$_2$ and MoS$_2$ on MgH$_2$ has been investigated in details.

## 2. Experimental section

### 2.1. Synthesis of a few layered WS$_2$

The bulk tungsten sulfide (WS$_2$) (99.80 %) powder was procured from the Alfa Aesar for the present investigation. For the preparation of few layered WS$_2$, WS$_2$ powder was dispersed in de-ionized water and sonicated it for 74 hours using ultrasonicator at 20 kHz frequency. The sonicated sample was then dried at 50 °C under a dynamic vacuum of order 10$^{-2}$ torr to form the few layered WS$_2$ powder. This preparation method can also be understood by the schematic given in Fig. 1.

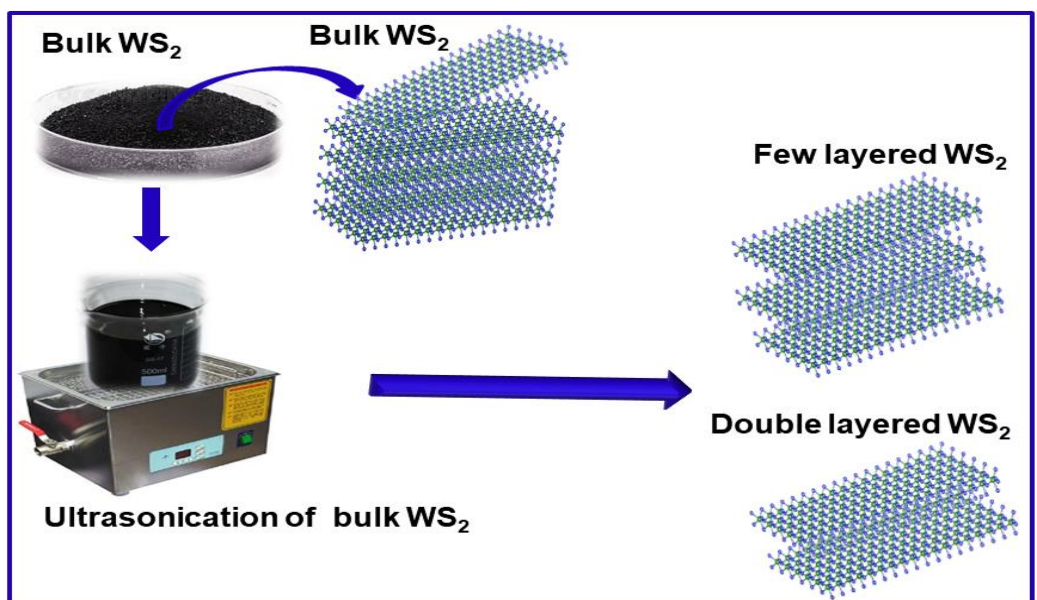

**Fig.1:** Schematic diagram for the synthesis of a few layers WS$_2$.

## 2.2. Synthesis of few-layer MoS$_2$

The Otto Chemica bulk molybdenum disulfide (MoS$_2$) (99 %) powder was used for the present investigation. MoS$_2$ powder was dispersed in de-ionized water and sonicate it for 74 hours using ultrasonicator at 20 kHz frequency to obtain the few-layered MoS$_2$. The sonicated sample was then dried at 50 °C under dynamic vacuum of order $10^{-2}$ torr to form the few layered MoS$_2$ powder. Fig. 2, shows the schematic diagram for preparation of few-layered MoS$_2$.

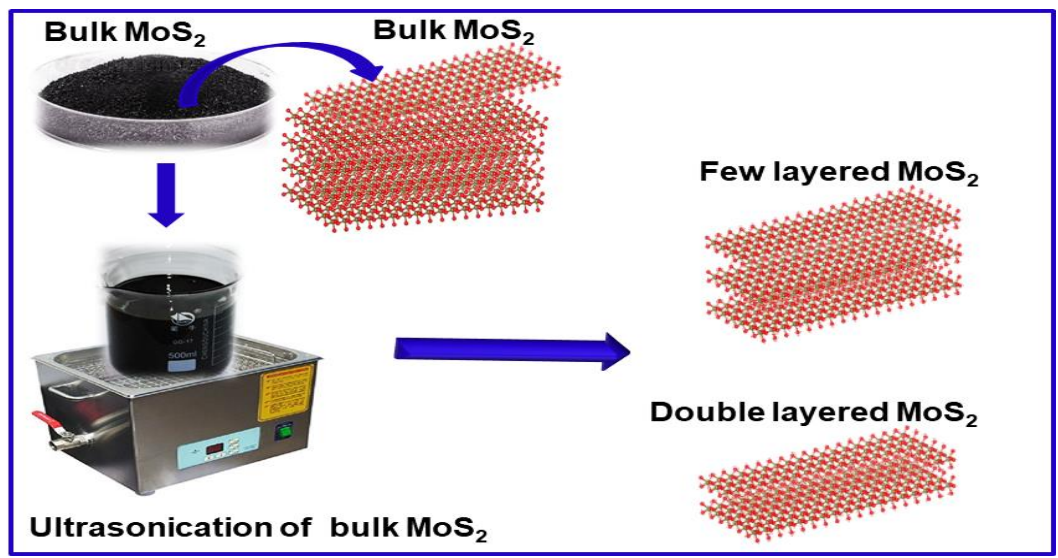

**Fig.2:** Schematic diagram for the synthesis of a few layers MoS$_2$.

## 2.3. Synthesis of MgH$_2$ catalyzed by WS$_2$, and MoS$_2$

The pure MgH$_2$ was procured from Fujifilm (Japan) (99.9%) for the present investigation. Mechanical ball-milling of MgH$_2$ with graphene at 180 rpm for 24 hours with a ball-to-powder ratio of 50:1 (by weight) using a planetary ball-miller (Retsch PM 400) was used to synthesize MgH$_2$ catalyzed by WS$_2$ (MgH$_2$-WS$_2$). To explore the optimum catalyst concentration for hydrogen sorption kinetics of Mg/MgH$_2$, we have synthesized a set of different catalyst concentrations (5, 10, 12 wt%) to catalyze MgH$_2$. For hydrogen sorption in Mg/MgH$_2$, 10 wt% catalysts were found to be optimal (in terms of desorption temperature and hydrogen storage capacity). The ball-miller vials were filled with 5 atm H$_2$ pressure to compensate for the loss of hydrogen from MgH$_2$ during milling. All the loading and unloading of the samples was done inside the N$_2$-filled glove box (MBRAUM MB10 compact) with O$_2$ and H$_2$O levels < 1 ppm. The synthesis of MgH$_2$ catalyzed by MoS$_2$ (MgH$_2$-MoS$_2$) was done using the same synthesis route as MgH$_2$-WS$_2$.

## 2.4. Characterization techniques

The structural characterization of prepared samples was carried out by XRD technique using Empyrean PANalytical X-ray diffractometer equipped with 2D detector with a Cu Kα beam (λ = 1.5415 Å) operated at 40 kV and 40 mA. The microstructural and selected area electron diffraction (SAED) analysis of as-prepared samples was carried out by TEM (Technai-20G$^2$) operating at the accelerating voltage of 200 kV. Perkin Elmer (Spectrum 100) spectrometer in transmission mode with attenuated total reflectance (ATR) sampling mode (wavenumber range 500–4000 cm$^{-1}$) was used to carry out FTIR spectroscopy. The Raman spectra have been acquired at -60 ℃ using Horiba-Jobin-Yvon LABRAM-HR800 spectrometer with diode LASER (532 nm). The desired thickness and surface topography of the prepared samples were examined by using solver next AFM in non-contact mode. The characterized samples then proceed for the hydrogen desorption and absorption using automated two-channel volumetric sieverts type apparatus. The temperature programmed desorption (TPD) was carried out with a heating rate of 5 ℃-

min$^{-1}$. The activation energy (E$_a$) study of prepared catalyzed samples has been done by using DSC (Perkin Elmer DSC 8000) with a heating rate of 15 $^{o}$C/min, 18 $^{o}$C/min, 21 $^{o}$C/min, and 24 $^{o}$C/min under nitrogen atmosphere (20 ml/min).

## 3. Results and discussion

### 3.1. Structural, microstructural, and spectroscopic characterization analysis

The structural characteristics of as-prepared samples have been examined using the XRD characterization. Fig. 3(a) shows the XRD pattern of pristine MgH$_2$, which matches well with the tetragonal MgH$_2$ with space group P42/mnm (136) and a=b= 4.516 Å, c = 3.020 Å (JCPDS no. 740934). Fig. 3(b) shows the XRD pattern of MoS$_2$, which matches well with the hexagonal structure of MoS$_2$ with space group P63/mmc(194) and a=b= 3.1602 Å, c = 12.294 Å (Joint Committee on Powder Diffraction Standards (JCPDS) no. 651951). The XRD pattern of as-prepared WS$_2$ is shown in Fig. 3(c), that matches well with the hexagonal structure of WS$_2$ with space group P63/mmc(194) and a=b= 3.1532 Å, c = 12.323 Å (JCPDS no. 841398). The usual diffraction pattern of MgH$_2$-MoS$_2$, and MgH$_2$-WS$_2$ are shown in Fig. 3(d-e), respectively, where besides the tetragonal phase of MgH$_2$, some peaks of WS$_2$ and MoS$_2$ are either suppressed or masked by the peaks of MgH$_2$. The diffraction peaks of WS$_2$ and MoS$_2$ are identified and labeled in the Fig. 3(d-e), respectively.

The different bands position, shapes, and relative intensities of Raman spectra give us essential information about the materials and stacking of layers, i.e., Raman spectroscopy can determine the layer thickness at the atomic level. The Raman spectra of as-prepared WS$_2$, and MoS$_2$ have shown in Fig. 4. In the case of MoS$_2$, the two Raman modes are appeared at ~ 345 cm$^{-1}$ and ~ 370 cm$^{-1}$ corresponds to E$^1_{2g}$ and A$_{1g}$ modes of vibrations (labeled in Fig. 4(b)). The indicated modes of MoS$_2$ have frequency difference of ~ 25 cm$^{-1}$, that means the MoS$_2$ as layered material with few layers of stacking (3-5 layers) [31,32]. The FWHM of A$_{1g}$ mode is ~ 7 cm$^{-1}$, which can also be referred to stacking a few layers of MoS$_2$ [33]. The Raman shifts at ~316 cm$^{-1}$ and 384 cm$^{-1}$ (shown in Fig. 4(a)) corresponds to the presence of E$^1_{2g}$ and A$_{1g}$ modes respectively in WS$_2$ sample. The intensity ratio of $E^1_{2g}$ and $A_{1g}$ modes was estimated $E^1_{2g}/A_{1g}$ i.e. = 1.26, which is higher than the intensity ratio of bulk $WS_2$ ($E^1_{2g}/A_{1g}$ = 0.47) and lower than the monolayer $WS_2$ ($E^1_{2g}/A_{1g}$ = 2.2) [34,35]. This calculated intensity ratio ($E^1_{2g}/A_{1g}$ = 1.26) is compatible with the range of 2-3 layers of $WS_2$.

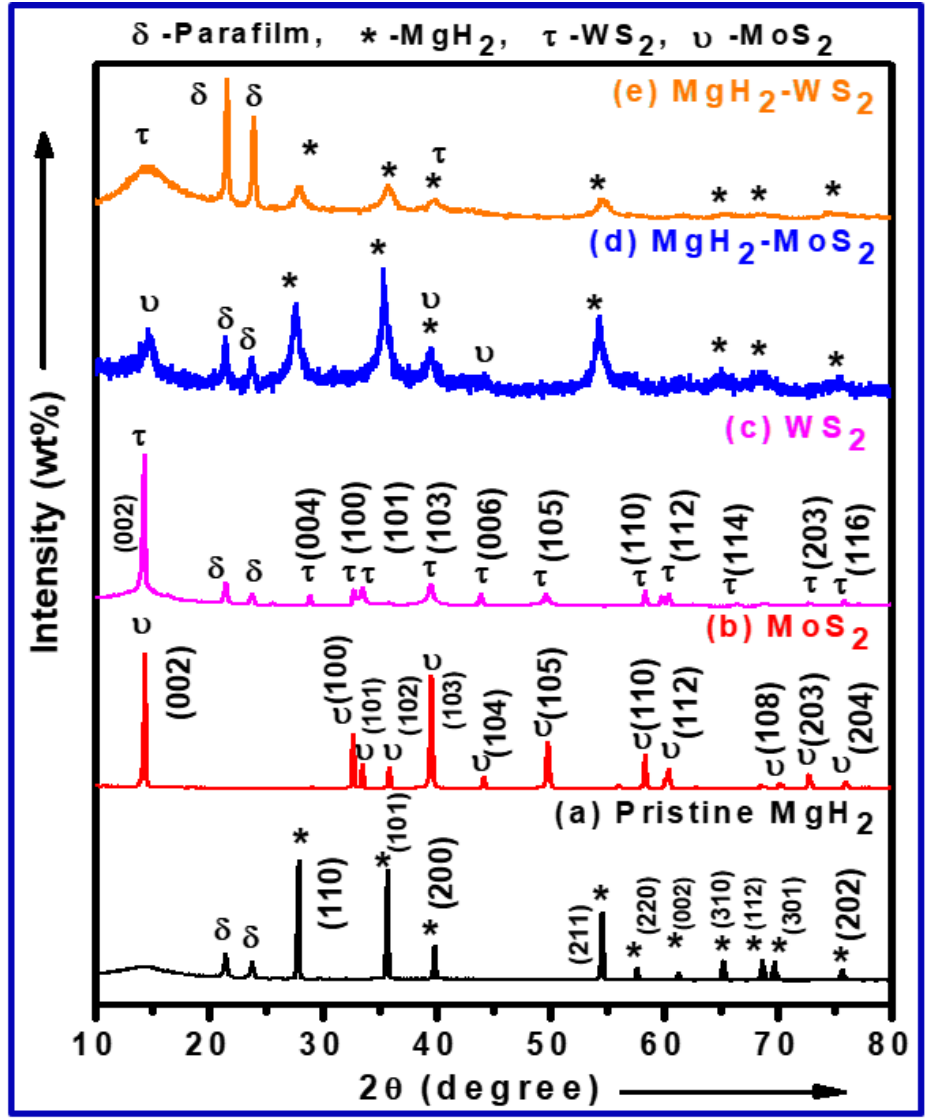

**Fig. 3:** XRD patterns of (a) Pristine $MgH_2$, (b) $MoS_2$, (c) $WS_2$, (d) $MgH_2$-$MoS_2$, and (e) $MgH_2$-$WS_2$.

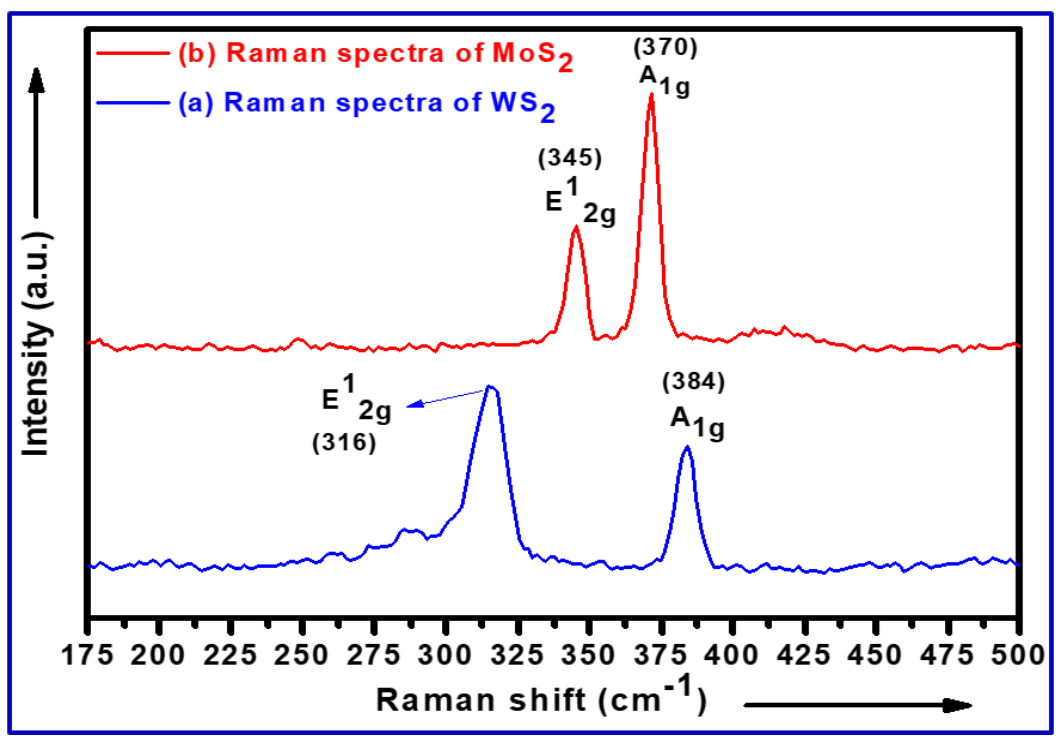

**Fig. 4:** Raman spectra of (a) WS₂ and (b) MoS₂.

The information about stacking layers in 2D layered materials (like WS₂ and MoS₂) can also be verified by AFM analysis. The surface topography and height profile of prepared MoS₂, and WS₂ were examined along the blue dotted line as shown in Fig. S1(a-b) (given in supporting information). The layered surface morphology along with height profile (shown in Fig. S1(a-a1)) shown the average thickness of MoS₂ is ~1.3 nm, that indicates the presence of ~2 layers of stacking in the MoS₂ sample [36,37]. The ~7-8 layers of stacking were present in the case of WS₂ (shown in Fig. S1(b-b1)) with a monolayer height of ~0.7 nm [38].

.

## 3.2 De/Re-hydrogenation kinetics of catalyzed MgH$_2$

To identify the optimal percentage of catalyst in MgH$_2$ with optimum temperature range where material performed promptly, we have characterized as-prepared samples for the temperature programmed desorption (TPD) analysis. The TPD curves of MgH$_2$-MoS$_2$ have seen in Fig. S2 (given in supporting information). The MgH$_2$-5%MoS$_2$, MgH$_2$-10%MoS$_2$, and MgH$_2$-12%MoS$_2$, starts releasing hydrogen at ~ 357 °C, ~ 330 °C, ~ 302 °C with ~ 6.41 wt%, ~ 6.00 wt%, ~ 4.88 wt% of hydrogen storage capacity respectively. On the other hand, MgH$_2$-5%WS$_2$, MgH$_2$-10%WS$_2$, and MgH$_2$-12%WS$_2$, starts releasing hydrogen at ~ 339 °C, ~ 277 °C, ~ 258 °C with ~ 6.54 wt%, ~ 5.95 wt%, ~ 5.14 wt% of hydrogen storage capacity respectively (shown in Fig. S3 in supporting information). Based on TPD analysis, the optimum catalyst concentration for catalyzing MgH$_2$ is 10 wt% for all catalysts.

After getting information about the optimum catalyst for MgH$_2$, we compared the TPD analysis of all optimum catalyzed samples with pristine MgH$_2$, as shown in Fig. 5. The TPD of pristine MgH$_2$ (shown in Fig. 5(a)) was then carried out to compare hydrogen storage properties with catalyzed samples. The pristine MgH$_2$ has an onset desorption temperature of 376 $^o$C with a total release of ~7.45 wt% storage capacity. The onset desorption temperature of MgH$_2$-MoS$_2$ (MgH$_2$-10%MoS$_2$) is ~ 330 $^o$C, and it desorbs ~ 6.00 wt% hydrogen while the desorption gets completed at 396 $^o$C (Fig. 5(b)). In the case of MgH$_2$-WS$_2$ (MgH$_2$-10%WS$_2$), it starts desorbing hydrogen at ~ 277 $^o$C with a storage capacity of 5.95 wt% (Fig. 5(c)).

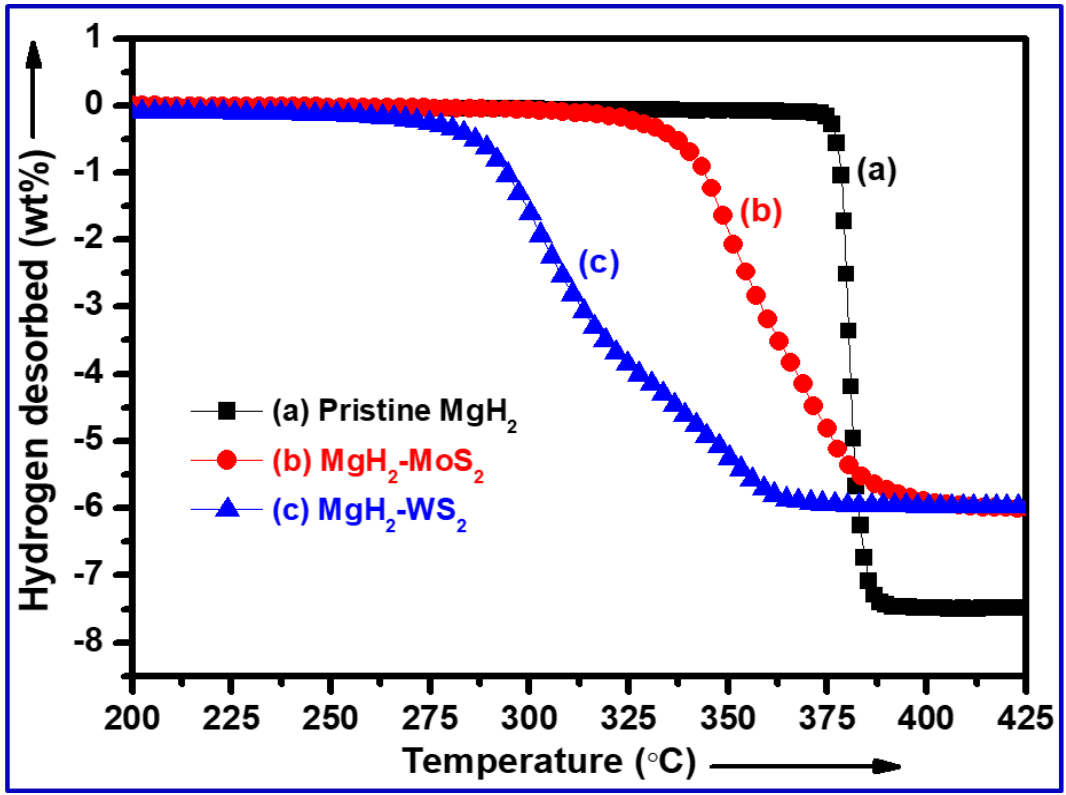

**Fig. 5:** Comparative TPD analysis of (a) Pristine $MgH_2$, (b) $MgH_2$-$MoS_2$ and (c) $MgH_2$-$WS_2$.

The desorbed samples then proceed for re/de-hydrogenation to check the cyclic stability and reversibility of catalyzed and pristine $MgH_2$. The re-hydrogenation kinetics was carried out at 300 °C under 13 atm hydrogen pressures, as shown in Fig. 6. It can be seen, the pristine $MgH_2$ absorbed ~1.16 wt% hydrogen in 1.2 minutes whereas $MgH_2$-$MoS_2$, $MgH_2$-$WS_2$ absorbed 4.60 wt%, 3.72 wt%, hydrogen, respectively, under similar conditions of temperature and pressure.

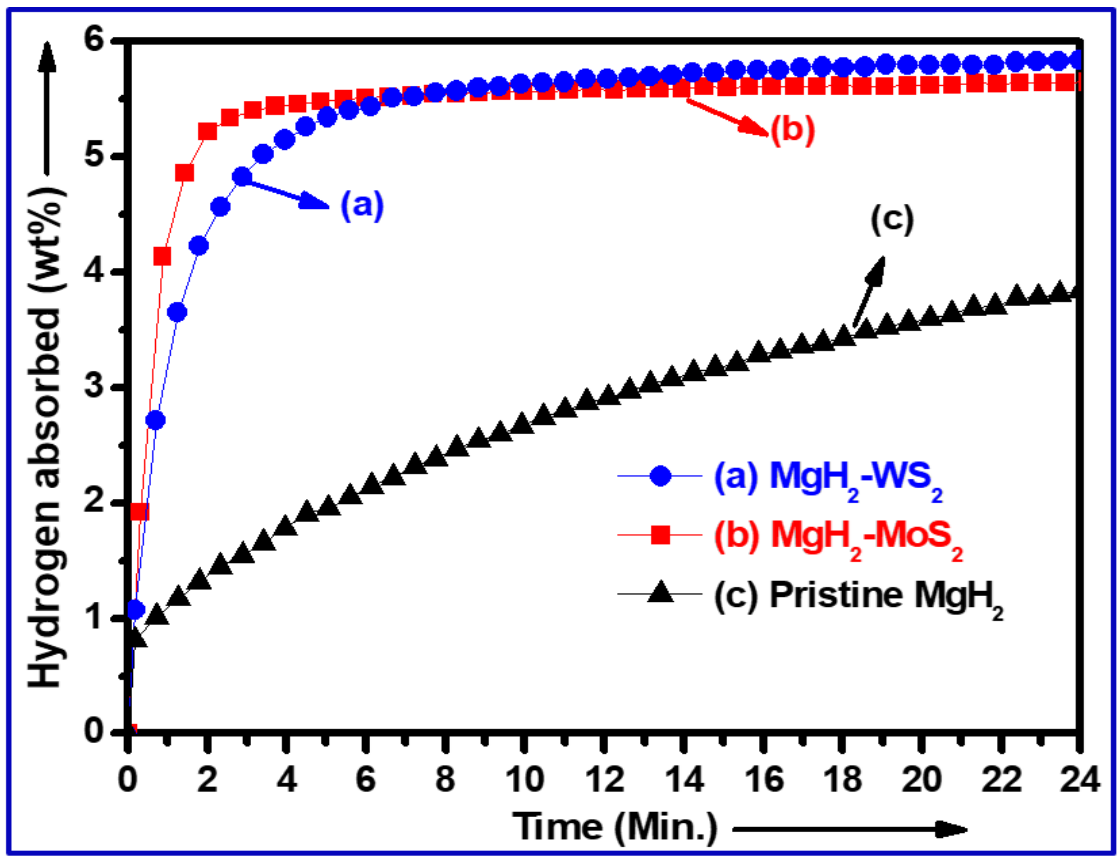

**Fig. 6:** Rehydrogenation kinetics curves at 300 °C under 13 atm H₂ pressure of (a) b) MgH₂-WS₂, (c) MgH₂-MoS₂ and (e) Pristine MgH₂.

The rehydrogenated samples were then dehydrogenated at 300 $^o$C under 1 atm hydrogen pressure. It can be seen clearly in Fig. 7, that the MgH₂-WS₂ sample releases 5.57 wt% hydrogen within 20 minutes while MgH₂-MoS₂ and pristine MgH₂ releasees 2.25 wt%, and 0.23 wt% of hydrogen under similar temperature and pressure conditions, which is 3.32 wt%, and 4.48 wt% more than pristine MgH₂, MgH₂-MoS₂, respectively. Based on the above re/de-hydrogenation kinetics study, it is clearly shown that WS₂ works as a superior catalyst to MoS₂ for catalyzing MgH₂. Therefore, in present study WS₂ is a prominent catalyst to catalyze MgH₂.

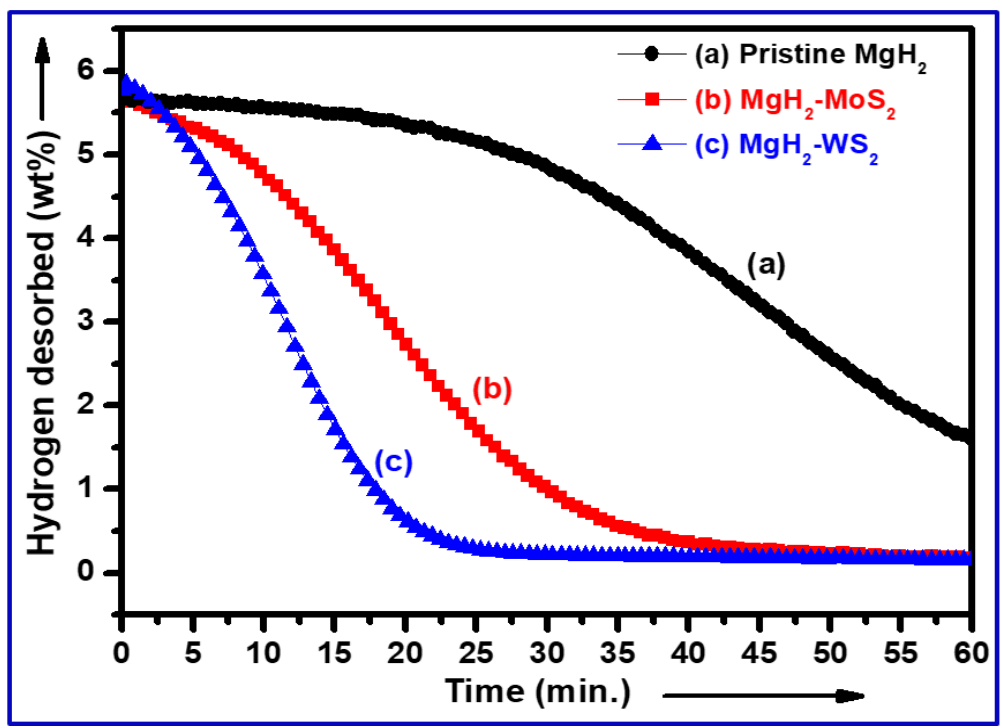

**Fig. 7:** Dehydrogenation kinetics curves at 300 °C under 1 atm H₂ pressure of (a) MgH₂-WS₂, (b) MgH₂-MoS₂, and (c) Pristine MgH₂.

### 3.3. Study of kinetics: Estimation of activation energy

The DSC was carried out to determine the hydrogen desorption activation energy barrier to convert MgH₂ into Mg. The DSC profile of MgH₂-MoS₂, MgH₂-WS₂, are shown in Figs. 8-9. In the case of MgH₂-WS₂, the peak desorption temperature found from DSC is ~ 380 °C, while the onset desorption temperature found from TPD is ~ 277 °C. There is a difference in desorption temperature in TPD (Fig. 5(c)) and DSC (Fig. 9(a)) curves due to the TPD being performed under vacuum with a temperature ramping rate of 5 °C/min while DSC was performed under N₂ atmosphere with a temperature ramping rate of 15 °C/min. For calculating the desorption activation energy, we have performed DSC with a set of the various rate of heating (15, 18, 21, 24 °C/min) and plotted the Kissinger curve by using the Kissinger equation[39] as given:

$$\ln(\beta/Tp^2) = \left(-\frac{E_a}{RT_p}\right) + \ln(k_O) \qquad (1)$$

Where $\beta$, $T_p$, and $E_a$ are the heating rate, corresponding peak desorption temperature, and activation energy, respectively. The slope of Kissinger plot ($\ln(\beta/T_p^2)$ vs. $1000/T_p^2$ plot) (Figs. 8-9) is used to calculate the desorption activation energy. The calculated activation energy for $MgH_2$-$MoS_2$, and $MgH_2$-$WS_2$ is 117.09 kJ/mol ($\pm$ 1.60 kJ/mol), and 104.00 kJ/mol ($\pm$ 2.74 kJ/mol) respectively. This activation energy indicates that ~104 kJ/mol energy is required to overcome the barrier to convert $MgH_2$ into Mg in the presence of a $WS_2$ catalyst. These calculated activation energies are significantly lower than the activation energy of pristine $MgH_2$ [3,40].

**Table 1:** Table for plateau pressures at corresponding temperatures, change in enthalpy, and activation energy of $MgH_2$-$MoS_2$, and $MgH_2$-$WS_2$.

| S.No. | Sample name | Plateaus pressure (atm) | Temperature ($^\circ$C) | Change in enthalpy (kJ/mol) | Activation energy (kJ/mol) |
|---|---|---|---|---|---|
| **1.** | $MgH_2$-$MoS_2$ | 1.03 | 272.62 | -78.33 | 117.09 |
| | | 2.03 | 292.28 | | |
| | | 3.77 | 313.28 | | |
| **2.** | $MgH_2$-$WS_2$ | 1.52 | 281.26 | -77.44 | 104.66 |
| | | 2.92 | 300.60 | | |
| | | 3.45 | 316.29 | | |

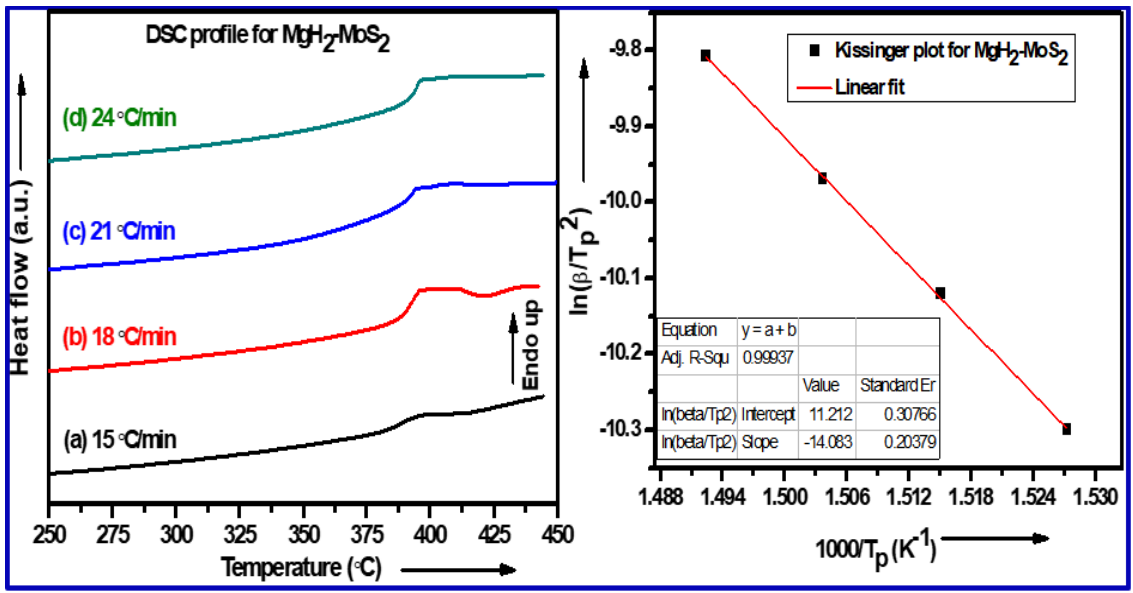

**Fig. 8:** (i) DSC profile for desorption of MgH₂-MoS₂ with the heating rate (a) 15 °C/min, (b) 18 °C/min, (c) 21 °C/min, (d) 24 °C/min, and (ii) corresponding Kissinger plot for evaluating the desorption activation energy of MgH₂-MoS₂.

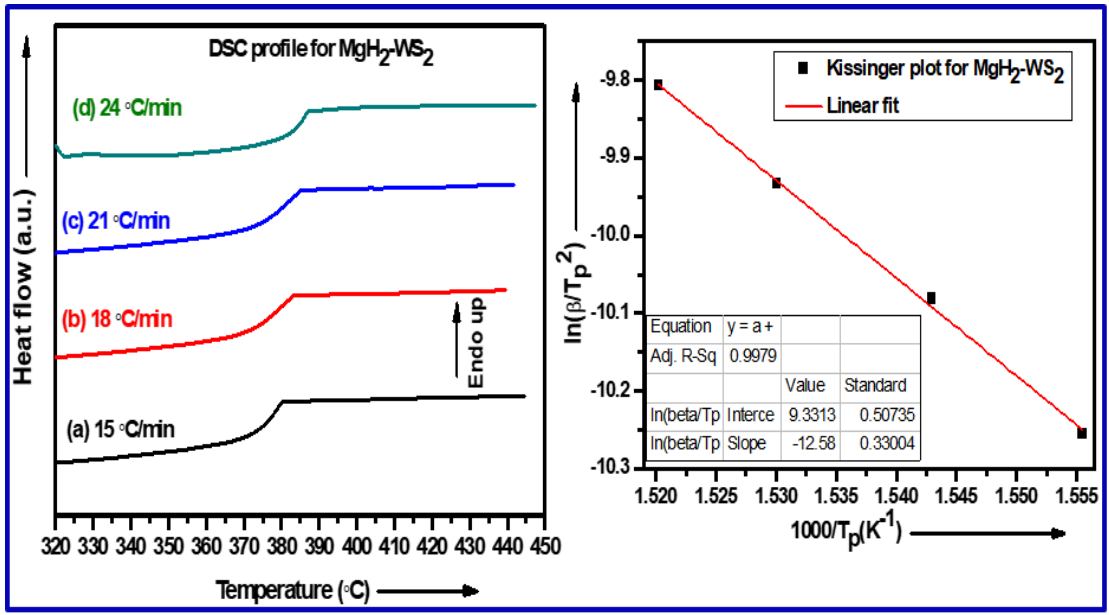

**Fig. 9:** (i) DSC profile for desorption of MgH₂-WS₂ with the heating rate (a) 15 °C/min, (b) 18 °C/min, (c) 21 °C/min, (d) 24 °C/min, and (ii) corresponding Kissinger plot for evaluating the desorption activation energy of MgH₂-WS₂.

### 3.4. Study of thermodynamics

After the kinetics and reversibility study, we have proceeded with the thermodynamic analysis of catalyzed MgH₂ for comparing the change in enthalpy and entropy of the system using well known Van't Hoff equation [41].

$$\ln P = (\Delta H / RT) - (\Delta S / R) \qquad (2)$$

Where P, $\Delta H$, R, T, and $\Delta S$ are the pressure, change in enthalpy, gas constant, absolute temperature, and change in entropy, respectively. The PCI isotherms (Figs. 10(i)-11(i) and Van't Hoff plots (Figs. 10(ii)-11(ii)) were used for the calculation of change in enthalpy of MgH₂-MoS₂ and MgH₂-WS₂, respectively. The calculated change in desorption enthalpy was found to be 78.33 kJ/mol (± 1.40 kJ/mol), and 77.44 kJ/mol (± 1.13 kJ/mol), for MgH₂-MoS₂ and MgH₂-WS₂ respectively. It is clear from the above estimation of change in enthalpy, that there is no significant enthalpy change in the presence of a catalyst. Thus MoS₂, WS₂ have not positively impacted the thermodynamic barrier of the MgH₂. The plateau pressure at corresponding temperatures, change in enthalpy, and activation energy has been tabulated in Table 1.

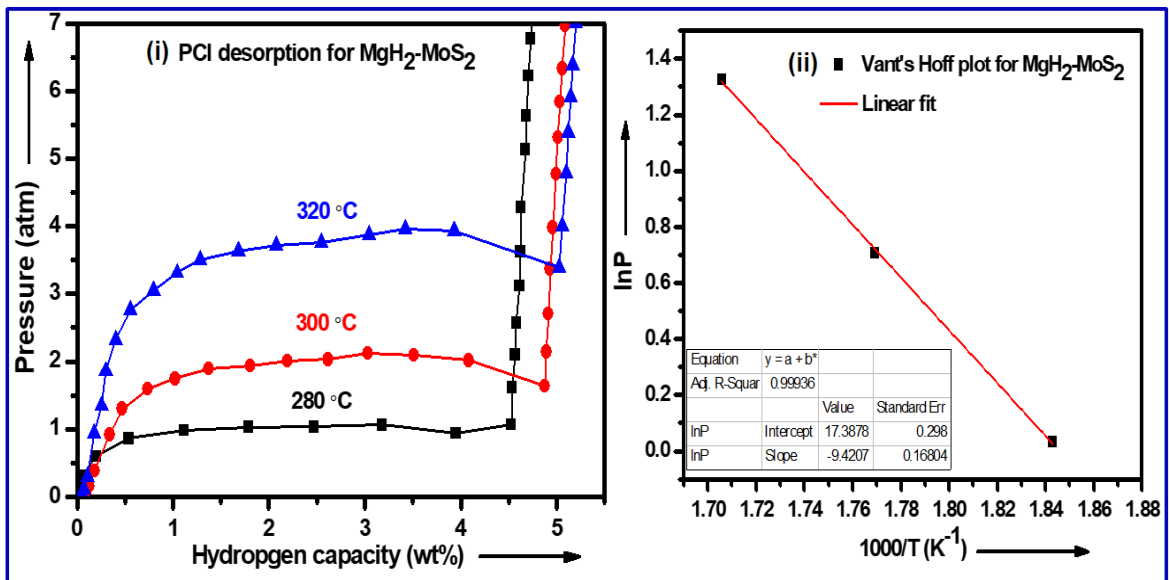

**Fig. 10:** (i) PCI desorption plots for MgH₂-MoS₂ at different temperatures and (ii) corresponding Van't Hoff plot for calculating the change in enthalpy

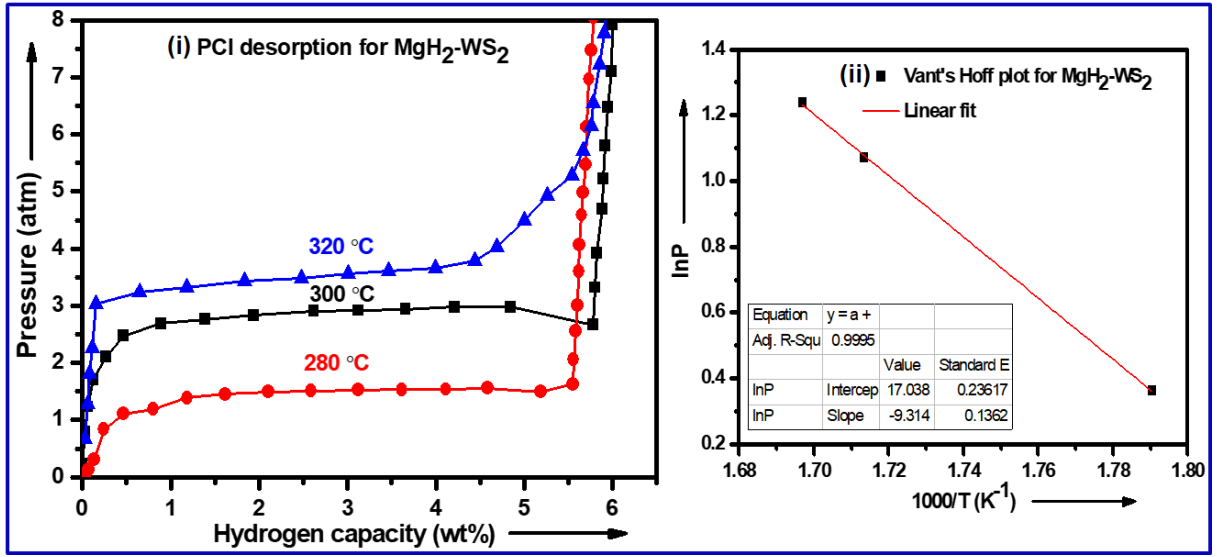

**Fig. 11:** (i) PCI desorption plots for MgH₂-WS₂ at different temperatures and (ii) corresponding Van't Hoff plot for calculating the change in enthalpy

### 3.5 Cyclic stability of catalyzed MgH₂

The WS₂ (optimum catalyst) plays a significant role in improving the kinetics of MgH₂. The cyclic stability is an essential characteristic of the hydride material (MgH₂) besides kinetic and thermodynamics, making it a worthy hydrogen storage material. Therefore, it is crucial to look at the cyclic stability of the catalyzed MgH₂ samples. We have performed 25 cycles of dehydrogenation (under 1 atm hydrogen pressure at 300 °C) and re-hydrogenation (under 13 atm hydrogen pressure at 300 °C) to ensure cyclic stability of catalyzed MgH₂. The cyclic stability curve of MgH₂-MoS₂ and MgH₂-WS₂ are shown in Fig. 12. From Fig. 12(a) MgH₂-MoS₂ shows the ~ 0.42 wt% (from 5.77 wt% to 5.35 wt%) degradation in hydrogen storage capacity during rehydrogenation and ~ 0.38 wt% (from 5.69 wt% to 5.31 wt%) in dehydrogenation. The MgH₂-WS₂ has the loss of hydrogen storage capacity ~ 0.3 wt% (from 5.80 wt% to 5.50 wt%) during re-hydrogenation and ~ 0.36 wt% (from 5.76 wt% to 5.40 wt%) during dehydrogenation. Thus, MgH₂-WS₂ has more substantial cyclic stability than MgH₂-MoS₂ under similar temperature and pressure conditions. The comparative study for hydrogen storage properties of different recently used 2D materials as the catalyst for MgH₂ is explored in Table 2.

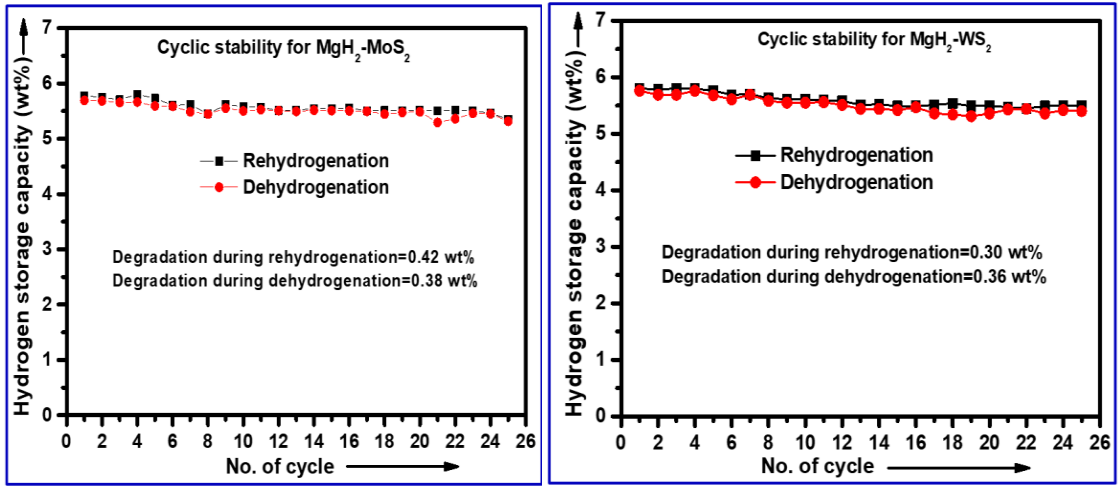

**Fig. 12:** Cyclic stability of (a) MgH₂-MoS₂ and (b) MgH₂-WS₂.

**Table 2:** Table for different 2D materials as the catalyst for hydrogen storage application.

| S. No. | Material | 2D- based catalyst | Hydrogen storage capacity (wt%) | Onset dehydrogenation temperature (°C) | Activation energy (kJ/mol) | Change in enthalpy (kJ/mol) | Ref. |
|---|---|---|---|---|---|---|---|
| 1. | Mg₆C₂N | C₂N | 6.79 | -- | -- | -- | [20] |
| 2. | MgH₂-LiAlH₄-Ti₃C₂ | Ti₃C₂ | 6.50 | 63.0 | 128.4 | 74.3 | [22] |
| 3. | MgH₂-Nb₄C₃Tₓ | Nb₄C₃Tₓ | 3.50 | 150.6 | 81.2 | -- | [21] |
| 4. | 1T'-MoS₂ | | 3.90 | -- | -- | -- | [42] |
| 5. | MgH₂-Gr | Graphene | 5.80 | 300.0 | -- | -- | [43] |
| **6.** | MgH$_2$-TiH$_2$@Gr | Graphene | 6.77 | 204.0 | 88.89 | 74.54 | [3] |
| | MgH$_2$-TiO$_2$@Gr | Graphene | 5.98 | 240.0 | 98.00 | 76.87 | |
| | MgH$_2$-Ti@Gr | Graphene | 5.70 | 235.0 | 103.03 | 75.65 | |
| **7.** | MgH$_2$-Gr | Graphene | 6.14 | 300.0 | 134.95 | 77.90 | [13] |
| **8.** | MgH$_2$-VS$_2$ | VS$_2$ | 6.51 | 242.0 | 98.10 | 76.83 | |
| **9.** | MgH$_2$-WS$_2$ | WS$_2$ | 5.95 | 277.0 | 104.66 | 77.44 | Present study |
| **10.** | MgH$_2$-MoS$_2$ | MoS$_2$ | 6.00 | 330.0 | 117.09 | 78.33 | Present study |

## 4. Conclusions

The catalytic effect of MoS$_2$, and WS$_2$ on MgH$_2$ was evaluated and compared. Based on the de/re-hydrogenation study, it is found that WS$_2$ works as an optimum catalyst over MoS$_2$ for MgH$_2$. The MgH$_2$-WS$_2$ has an onset de-hydrogenation ~277 $^o$C with a hydrogen storage capacity of 5.95 wt%. The MgH$_2$-WS$_2$ absorbed hydrogen ~ 3.72 wt% within 1.3 minutes at 300 $^o$C under 13 atm hydrogen pressure and it desorbed ~5.57 wt% within 20 minutes at 300 $^o$C under 1 atm hydrogen pressure. The MgH$_2$-WS$_2$ shows a minimum degradation of hydrogen storage capacity ~ 0.3 wt% upto 25 cycles which shows a better cyclic stability than cyclic stability of MgH$_2$-MoS$_2$ (~ 0.4 wt% loss in hydrogen storage capacity). MgH$_2$-WS$_2$ also shows the lower reaction activation energy ~117 kJ/mol as compare to other catalyzed and uncatalyzed samples. On the other hand, these catalysts (WS$_2$ and MoS$_2$) do not have any impact on the thermodynamical parameters that is change in enthalpy. This study opens a new era to further applications of 2D layered materials for various applications like template materials.

## Acknowledgments

We gratefully accept funding assistance from the Department of Science and Technology (DST), New Delhi, India. The Council of Scientific and Industrial Research (CSIR), New Delhi, India, has awarded the author (S.K.V.) a CSIR-Senior Research Fellowship (Award No. 09/013(0872)/2019-EMR-I), for which the author is grateful.

## Conflict of Interest Declaration

There are no conflicts of interest among the authors.